\documentclass[aps,10pt,twocolumn,showpacs]{revtex4}

\usepackage{amsfonts}

\usepackage{amsmath}

\usepackage{amssymb}

\usepackage{graphicx}

\begin{document}

\title[Short title for running header]{Electric Field Induced Symmetry Breaking\\ of Angular Momentum Distribution in Atoms}

\author{Marcis Auzinsh}

\email{mauzins@latnet.lv}


\affiliation{University of Latvia, Department of Physics and Mathematics,
  Rainis Blvd 19, Riga LV-1586, Latvia}

\author{Kaspars Blushs}

\affiliation{University of Latvia, Department of Physics and Mathematics,
  Rainis Blvd 19, Riga LV-1586, Latvia}

\author{Ruvin Ferber}

\affiliation{University of Latvia, Department of Physics and Mathematics,
  Rainis Blvd 19, Riga LV-1586, Latvia}

\author{Florian Gahbauer}

\affiliation{University of Latvia, Department of Physics and Mathematics,
  Rainis Blvd 19, Riga LV-1586, Latvia}

\author{Andrey Jarmola}

\affiliation{University of Latvia, Department of Physics and Mathematics,
  Rainis Blvd 19, Riga LV-1586, Latvia}

\author{Maris Tamanis}

\affiliation{University of Latvia, Department of Physics and Mathematics,
  Rainis Blvd 19, Riga LV-1586, Latvia}



\begin{abstract}
We report the experimental observation of alignment to orientation conversion
in the 7D$_{3/2}$ and 9D$_{3/2}$ states of Cs in the presence of an 
external dc electric field, and without the influence of magnetic fields or
atomic collisions.  Initial alignment of angular momentum states was created
by two-step excitation with linearly polarized laser radiation.  
The appearance of transverse orientation of angular momentum was 
confirmed by the  observation of circularly polarized light.  We present
experimentally measured signals and compare them with the results of a
detailed theoretical model based on the optical Bloch equations.

\end{abstract}

\date{\today }

\maketitle

Atomic physics experiments are able to place sensitive limits on the
permanent electric dipole moment (EDM) of the electron because highly
polarizable atoms can generate internal electric fields that are orders of 
magnitude higher than can be achieved for free particles~\cite{Khr97,San2001}. 
Such limits on the 
electron EDM are of great interest to fundamental physics because a permanent 
EDM of a fundamental particle implies CP violation as long as
the CPT theorem is assumed to hold.  Such a CP violation would be a signature
of new physics beyond the standard model~\cite{Gin2004}. The most
sensitive upper limit to date on the electron EDM
has been achieved by searching for a small precession around an
external electric field of the angular momentum distribution of an ensemble
of atoms~\cite{Reg2002}. Various sophisticated experimental techniques
prevent angular momentum precession caused by mechanisms other than an EDM
from contaminating the signal.  One such phenomenon,
known as alignment to orientation conversion (AOC), is known to deform the
atomic and molecular angular momentum distributions under
the influence of combined electric and magnetic fields or
collisions.  Moreover, it has been shown theoretically \cite{Auz94} that AOC
can be induced by a purely electric field without the need for magnetic
fields or collisions whenever the initial alignment is not exactly 
perpendicular or parallel to the external electric field. 
In this letter, we report the experimental observation of
AOC in the presence of only an electric field.  Our measured signals
are described accurately by a detailed theoretical model, which could be of
interest in studying potential backgrounds for electron 
EDM searches in atomic or molecular systems.

Angular momentum \emph{alignment} means that the population of atomic 
sublevels $m_J$ varies with $|m_J|$, but $+m_J$ and $-m_J$ levels are
populated equally.  When the $+m_J$ and $-m_J$ levels are unequally populated,
the ensemble is \emph{oriented}.
One can describe the anisotropic spatial distribution of 
angular momenta {\bf J} in an ensemble
by an atomic density matrix $\rho$ \cite{Blu96}.
An ensemble is described as \emph{aligned} if the distribution of
angular momenta contains a net electric quadrupole moment or \emph{oriented}
if there is a net magnetic dipole moment.  Orientation can be classified
as longitudinal or transverse, with reference to the symmetry axis.  The 
breaking of the reflection symmetry of an initially aligned population of
atoms by converting the alignment into transverse orientation has been a 
subject of theoretical and experimental investigation since the 1960s, 
beginning with the work of \cite{Fan64,Reb68,Lom67,Lom69}.

The very first theoretical works suggested that AOC may 
be induced in atoms by a magnetic field gradient~\cite{Fan64} or by 
anisotropic collisions in which the angle between the collision axis 
and the alignment axis differs from 0 and $\pi/2$ ~\cite{Reb68,Lom67}.   
These latter predictions were confirmed by~\cite{Cha77,Man81}.  
Orthogonal static electric and  magnetic fields were used to study Ba and 
Cs~\cite{Hil94}, while magnetic fields only were used to study AOC in
Rb~\cite{Kra79,Aln2001}, and Na~\cite{Han91}.  When a strong pumping laser
field excites atoms in the presence of a magnetic field, the interaction of 
the dynamic Stark effect with the magnetic field was shown to lead to AOC, 
first in Hg \cite{Coh69} and then in other elements~\cite{Bud2000,Kun2002}.
In particular, a sizeable degree of circular
polarization (c.a. 15\%), caused by AOC, was observed in the $(5d6p)^1P$ state 
of Ba in a femtosecond regime due to the combined effects of a static
magnetic field and an electric field from a pulsed laser~\cite{Kun2002}.
Lombardi \emph{et al.}~\cite{Lom68,Lom69} measured the appearance of 
orientation in the 4$^{1}$D$_{2}$ state of helium using a capacitative 
electrodeless helium discharge.  In this case, collisions with electrons
served as a source of initial alignment, which was then transformed into
orientation by a combination of static electric and magnetic fields.  
Similar experiments were performed by Elbel \emph{et al.}~\cite{Elb90}.  
Since then, there have been experiments that have shown how AOC can be caused 
in atomic systems by magnetic field gradients, collisions, and combined 
electric and magnetic fields.   However, no experimental studies of AOC in 
atoms in a purely electric field have been reported.

It has been predicted that the quadratic Stark effect in the presence of
an electric field only could cause AOC \cite{Auz95,Auz97} without any need for
magnetic fields or collisions.  
In general, linearly polarized light is expected to produce alignment, but not
orientation.  However, initial alignment produced by linearly polarized
light can be converted into transverse orientation when the aligned atoms  are
placed in a perturbing
electric field that makes an angle with the initial alignment axis different 
from 0 or $\pi/2$ and the quadratic Stark effect causes levels with 
$\Delta m = \pm 1$ to cross.

In this letter we confirm these predictions
with quantitative measurements, which we compare with the results of a
simulation.

\begin{figure}[tbp]
\centering
\includegraphics[width=0.4\textwidth]{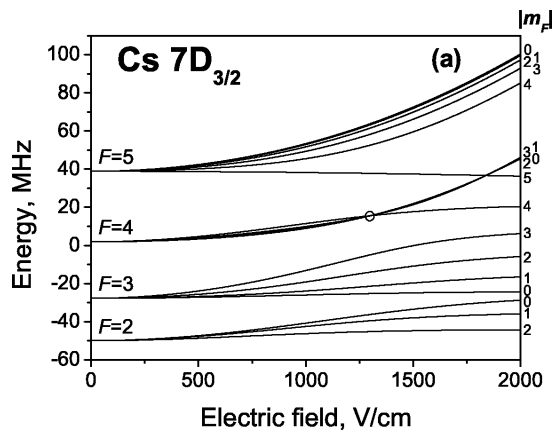}
\vspace{0.5cm}
\includegraphics[width=0.4\textwidth]{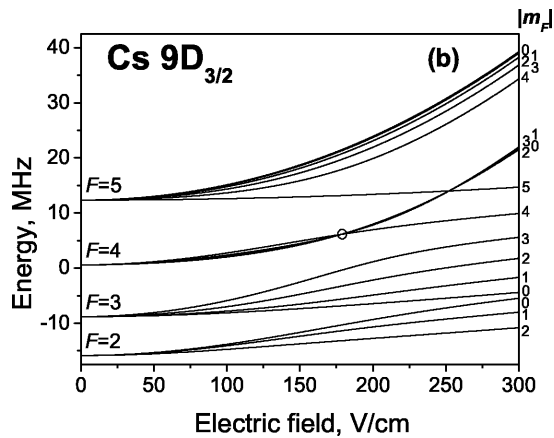}
\caption{Hyperfine level splitting diagram in an external electric field for
  the (a) 7D$_{3/2}$ and (b) 9D$_{3/2}$ states of Cs.}
\label{lc}
\end{figure}

  In our experiment we studied AOC in the 7D$_{3/2}$ and 9D$_{3/2}$ states of 
Cs.   Figure~\ref{lc} shows 
the hyperfine level splitting in an external electric field for these states,
which are calculated by 
diagonalizing in an uncoupled basis the Hamiltonian, which includes the
hyperfine and Stark terms~\cite{Ale93}.  To compute this Hamiltonian, we use 
the following atomic constants: 
hyperfine constant $A$=7.4(2) MHz~\cite{Ari77},
scalar polarizability $\alpha_0=-6.0(8)\times 10^{4}$\cite{Wes87}, and 
tensor polarizability $\alpha _{2}=7.45(20)\times 10^{4}$~\cite{Auz05} 
in $a_{0}^{3}$
units for the 7D$_{3/2}$ state; and $A$=2.35(4) MHz~\cite{Ari77}, 
$\alpha_{0}=-1.45(12)\times 10^{6}$\cite{Fre77}, and 
$\alpha _{2}=1.183(35)\times 10^{6}$~\cite{Auz05} for the 9D$_{3/2}$ state.

\begin{figure}[tbp]
\includegraphics[width=0.375\textwidth]{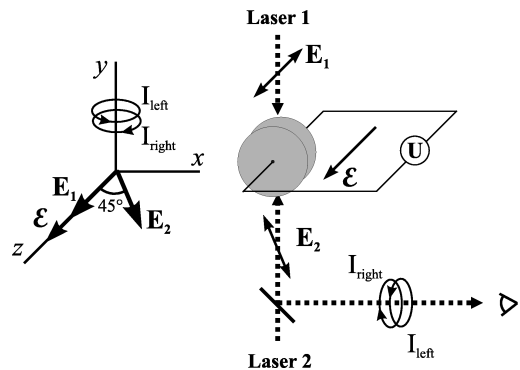}
\caption{Experimental geometry.}
\label{exp}
\end{figure}

The circle in each diagram marks a crossing of
magnetic sublevels with $\Delta m = \pm 1$.  
Two sublevels with $\Delta m = \pm 1$ can be simultaneously excited with
linearly polarized light when they are degenerate in energy and  
there is a non-zero
angle between the polarization vector of the exciting radiation and the
quantization axis, i.e., the electric field axis.
The relative orientations of the electric field {\bf $\mathcal{E}$} and the 
laser polarization vectors {\bf E$_1$} and {\bf E$_2$} and the 
laser induced fluorescence (LIF) observation
direction are shown in Fig.~\ref{exp}.
In this setup, the appearance of transverse 
orientation, that is, orientation perpendicular to the quantization axis,
can be detected by observing circularly polarized light in a 
direction perpendicular to {\bf $\mathcal{E}$} 
as well as {\bf E$_1$} and {\bf E$_2$}.

\begin{figure}[tbp]
\includegraphics[width=0.375\textwidth]{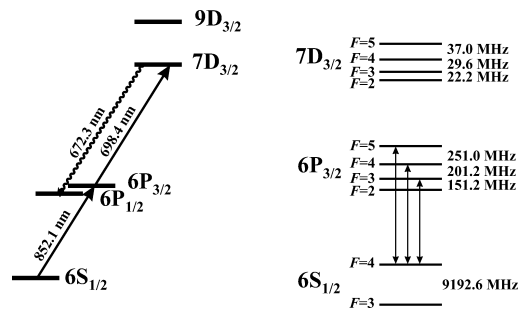}
\caption{Cs energy-levels and excitation-observation scheme.}
\label{excitation}
\end{figure}

We make measurements on cesium vapor contained in a glass cell at room 
temperature. In order to reach the 7D$_{3/2}$ and 9D$_{3/2}$ states of cesium 
we use two-step laser excitation with counterpropagating laser beams
(see Fig.~\ref{excitation}). In the first 
step atoms are excited to the 6P$_{3/2}$ state by a diode laser 
(LD-0850-100sm laser diode), which is linearly polarized 
along the external dc electric field {\bf$\mathcal{E}$} direction 
({\bf $\mathcal{E}$} $\parallel$ {\bf E$_1$}). 
In the second step we use either a
diode laser (Hitachi HL6738MG laser diode) to induce the 
6P$_{3/2} \rightarrow$ 7D$_{3/2}$ 
transition, or a Coherent CR699-21 ring dye laser with Rodamin 6G dye to 
induce the 6P$_{3/2} \rightarrow$ 9D$_{3/2}$ transition. 
The polarization vector of the second laser {\bf E$_2$} 
makes an angle of $\pi/4$ with respect to that of the first {\bf E$_1$} and
the electric field {\bf $\mathcal{E}$} (see Fig~\ref{exp}). 

An electric field is applied via polished stainless steel Stark 
electrodes of 25 mm diameter located inside the cell and separated by a 5 mm 
gap.
The LIF at the $n$D$_{3/2} \rightarrow$ 
6P$_{1/2}$ transition is observed collinearly to the laser beams 
with the help of a pierced mirror.  
Before entering a monochromator, 
the LIF passes through a 
$\lambda/4$ plate, which converts circularly polarized light into linearly 
polarized light, allowing us to measure the degree of circularity $C$ defined 
as

\begin{equation}
C=\frac{I(E_{right})-I(E_{left})}{I(E_{right})+I(E_{left})},
\end{equation}
where $I(E_{right})$ and $I(E_{left})$ are the intensities of the right and 
left circularly polarized LIF.

The signal is detected by a photomultiplier tube in photon counting mode and 
recorded on a PC together with the electrode voltage. The voltage is applied 
in discrete steps. During each step, the number of photons is counted 
for $E_{right}$ and $E_{left}$.

\begin{figure}[tbp]
\includegraphics[width=0.4\textwidth]{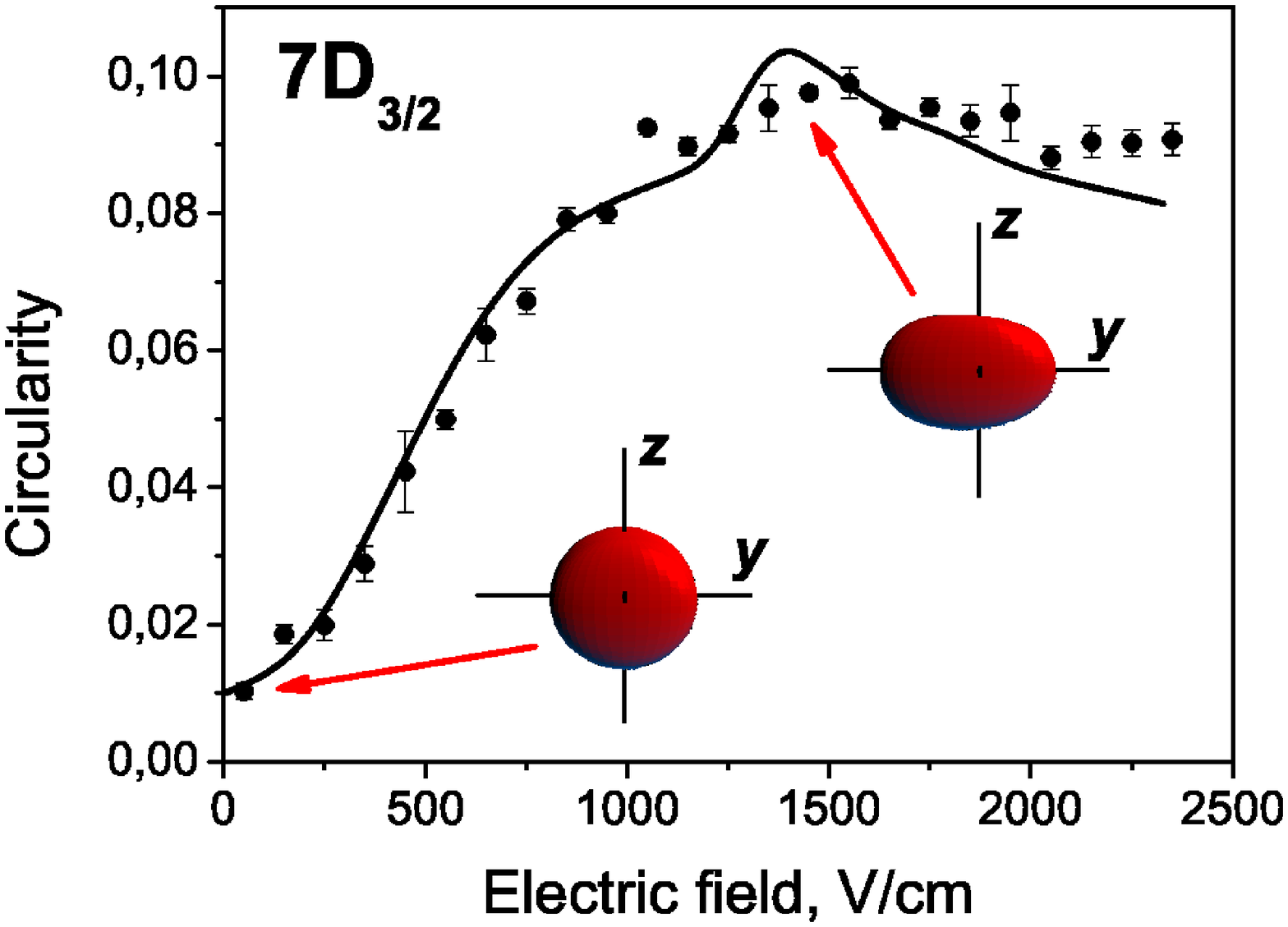}
\vspace{0.5 cm}
\includegraphics[width=0.4\textwidth]{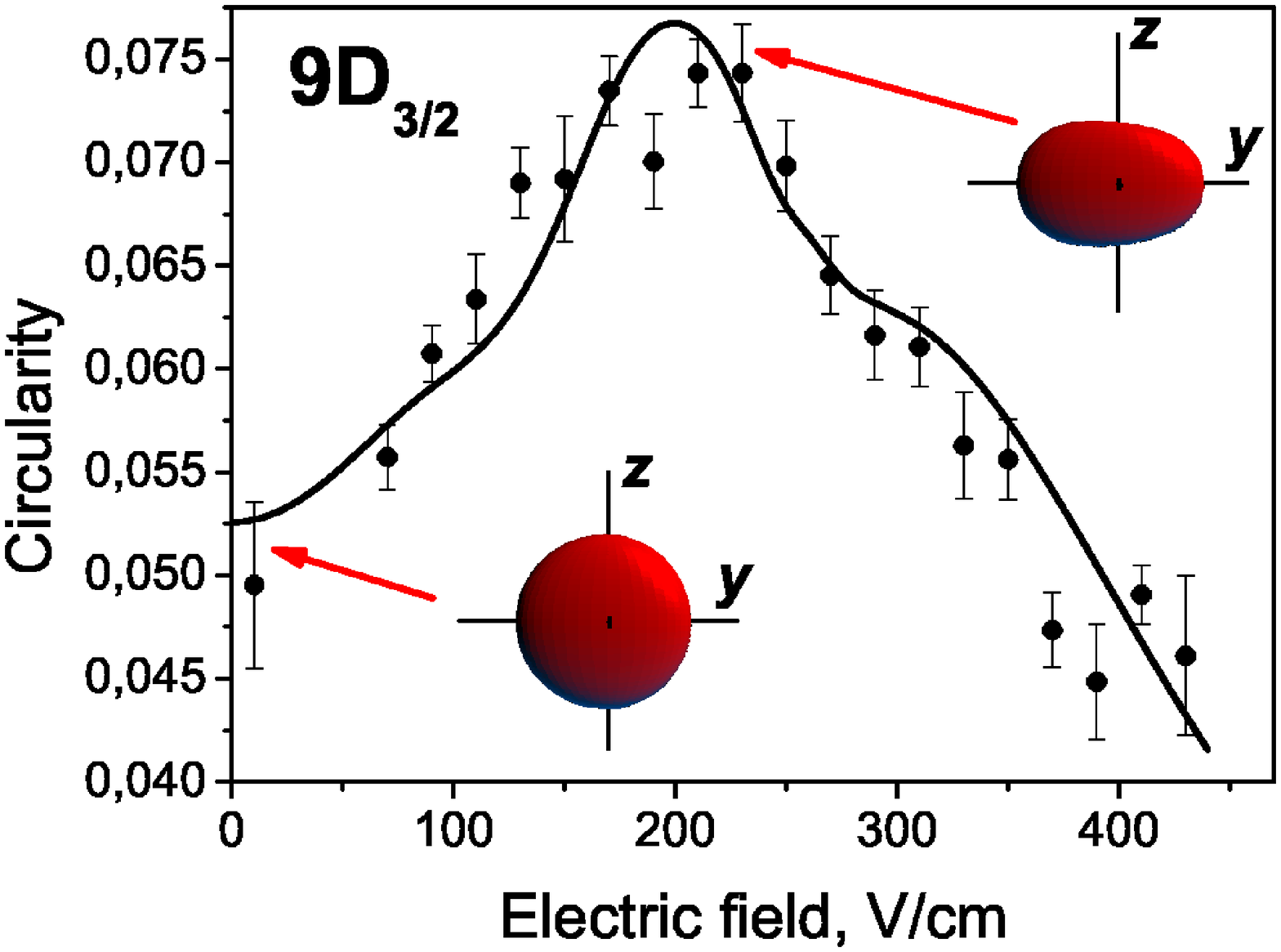}
\caption{Degree of LIF circularity $C$ as a function of electric field for the
  7D$_{3/2}$ and 9D$_{3/2}$ states of Cs. Points: experiment. Solid line: simulation.  Insets depict the atomic angular momentum distribution.}
\label{aoc}
\end{figure}

The results of our measurements are shown as dots in in Fig.~\ref{aoc}.   
The circularity $C$ reaches a maximum near the $\Delta m = \pm 1$ 
crossing (circled in Fig. 1).  A value of $C$ as high as 10\% is observed in 
the case of the 7D$_{3/2}$ state.  A small orientation appears at zero 
electric field because the linewidth of the second laser is sufficiently 
broad to excite coherently magnetic sublevels belonging to different $F$ 
states.  This effect is more prounounced in the 9D$_{3/2}$ state because
of the smaller hyperfine level splittings. 
Solid lines show the result of simulations.  Our theoretical
treatment builds on earlier
models developed to solve the rate equations for Zeeman coherences for
single-step~\cite{Blu2004} and double-step~\cite{Auz05} laser excitation
of atoms.

The model assumes that the atoms move classically and are excited by the
laser radiation at the internal transitions. In this case, the internal
dynamics of the atom can be described by a semiclassical atomic density
matrix $\rho $ parametrized by the classical coordinates of the atomic
center of mass. To obtain the density matrix, we must solve the optical
Bloch equations (see, for example, Ref.~\cite{Ste84})

\begin{equation}
i\hbar \frac{\partial \rho }{\partial t}=\left[ \widehat{H},\widetilde{\rho }
\right] +i\hbar \widehat{R}\rho .
\end{equation}%
The Hamiltonian $\widehat{H}$ includes the unperturbed Hamiltonian for the
hyperfine interaction, the dipole interaction operator, and the 
dc Stark operator.  The operator $\widehat{R}$ describes
relaxation and includes both spontaneous emission and transit relaxation.
We assume that the density of atoms is sufficiently low so that different
velocity groups of thermally moving atoms do not interact.
The model also accounts for nonlinear effects such as the ac 
Stark effect even though they are small in our experiment because 
the laser power was low.

We apply the rotating wave
approximation for multilevel systems \cite{Ari96} to eliminate oscillations
with optical frequencies. Since the time scale of the experiment is large
compared to the characteristic phase fluctuation time of the lasers, we take
a statistical average over the fluctuating phases~\cite{Blu2004}. We assume
that both lasers are uncorrelated, so that we can apply the
\textquotedblleft decorrelation approximation"~\cite{Kam76,Blu2004} to
obtain a system of equations, which, when solved, yields the density matrix
of the $n$D$_{3/2}$ state, from which the fluorescence intensities in each
polarization are obtained.

We obtained the values of those simulation parameters that could not be 
controlled precisely in the experiment by fitting simulations 
and measurements of level crossing signals of the same (7,9)D$_{3/2}$ Cs 
states, which were measured under the same experimental 
conditions~\cite{Auz05}.  
These parameters included
the spectral widths of the laser radiation and the
detuning with respect to the exact hyperfine transition frequencies.

  The excellent agreement between experiment and theory (see Fig.~\ref{aoc}) 
demonstrates the
validity of our theoretical approach. 

To convey an intuitive understanding of the AOC effect, Fig.~\ref{aoc}
contain inserts which help to visualize the angular
momentum distributions at zero electric field, and at the electric field
value for which the degree of orientation was at a maximum.  The angular 
momentum distribution is visualized as a surface whose distance from the 
origin in proportional to the probability that the angular momentum of an
atom in the ensemble points toward that point on the surface.  This probability
is computed from the components of the density matrix 
$\rho_{mm^{\prime}}$~\cite{Auz97,Roc2001}.  
The theoretical model involves computing the
density matrix $\rho_{MM^{\prime}}^{FF^{\prime}}$ for the manifold of 
hyperfine levels.  This matrix describes the population of certain $F$ states, 
as well
as the Zeeman coherences inside those states and between different $F$ states.
Since we do not resolve spectrally the $n$D state hyperfine components
that are separated by $\sim 10$ MHz, it is convenient in the final analysis 
to pass to the overall density matrix for a fine structure level.  To this 
end, we need to contract the fine structure density matrix over the nuclear
spin $I$ projections $\mu,\mu^{\prime}$ and sum over the hyperfine structure
components with the expansion coefficients of the fine structure wave
functions over the hyperfine structure wave functions

\begin{align}
\rho _{mm^{\prime }}& =\sum_{\mu }\rho _{m\mu m^{\prime }\mu ^{\prime
}}\delta _{\mu \mu ^{\prime }}=  \notag \\
& =\sum_{FF^{\prime }MM^{\prime }}\rho _{MM^{\prime }}^{FF^{\prime
}}\sum_{\mu }C_{JmI\mu }^{FM}C_{Jm^{\prime }I\mu }^{F^{\prime }M^{\prime }}.
\end{align}
In the absence of the external field these expansion coefficients are
the Clebsch-Gordan coefficients $C_{J_mI_{\mu}}^{FM}$.  If the
electric field is present, the Clebsch-Gordan coefficients must be 
replaced by the expansion coefficients that can be obtained by
diagonalizing the Stark Hamiltonian.

In summary, we have observed AOC produced by the quadratic Stark effect
in an external electric field without the need of any magnetic fields or
collisions.  With the light {\bf E} vectors and the external dc 
electric field {\bf $\mathcal{E}$} forming an angle of $\pi/4$, 
transverse orientation perpendicular to the {\bf E$\mathcal{E}$}-plane
appeared, and gave rise to LIF signals with a degree of circularity up
to 10\%.  In order to search for the electron EDM in atomic or molecular
systems, it is often crucial to ensure that the dc electric field,
the light polarization vectors, and the residual magnetic field are
exactly parallel or orthogonal to each other.  As experiments push the
limit even lower, this requirement will be hard to achieve with the
necessary accuracy, and so AOC could introduce a possible background.  Our
investigation shows that AOC can be understood very well and corrections
could be made, if necessary.

We thank Janis Alnis for assistance with the diode lasers, Robert Kalendarev
for preparing the cesium cell, and Dmitry Budker for helpful discussions. 
This work was supported by the NATO SfP 978029 Optical Field Mapping grant, 
and the Latvian State Research 
Programme funding grant 1-23/50.  K.B., F.G, and A.J. gratefully acknowledge 
support from the European Social Fund.

\bibliographystyle{apsrev}
\bibliography{cs_aoc_2}

\end{document}